\documentclass[
]{ceurart}
\pdfoutput=1
\usepackage{hyperref}
\hypersetup{
  pdfinfo={
    Title={Your Title Here},
    Author={Your Name Here},
    Subject={If you want to put something here, do so},
    Keywords={Add some keywords if you feel so inclined}
  }
}
\usepackage{pdfpages}

\usepackage{xcolor}

\usepackage{todonotes}

\usepackage{multirow}
\usepackage{wrapfig}
\usepackage{subfigure}
\usepackage{listings}
\usepackage{amsmath,amsthm,amssymb}
\usepackage{etoolbox}
\usepackage{array}
\usepackage{tikz}
\usetikzlibrary{shapes,arrows,arrows.meta, positioning,shapes.misc,decorations.markings}
\usepackage{adjustbox}
\usepackage{hyperref}
\usepackage{booktabs}
\usepackage{lipsum,multicol}
\usepackage[font=small,skip=4pt]{caption}
\usepackage{enumitem}
\usepackage{caption}
\usepackage{algorithm}
\usepackage[noend]{algpseudocode}

\DeclareMathOperator{\coocur}{co\_occur}
\DeclareMathOperator{\len}{len}
\DeclareMathOperator{\edist}{edit\_dist}
\DeclareMathOperator{\escore}{edit\_score}
\DeclareMathOperator{\queue}{Queue}
\DeclareMathOperator{\enqueue}{enqueue}
\DeclareMathOperator{\dequeue}{dequeue}
\DeclareMathOperator{\isempty}{is\_empty}
\DeclareMathOperator{\children}{children}
\DeclareMathOperator{\resolvepair}{resolve\_alias\_pair}

\begin{document}

\copyrightyear{2023}
\copyrightclause{Copyright for this paper by its authors. Use permitted under Creative Commons License Attribution 4.0 International (CC BY 4.0).}

\conference{CAMLIS'23: Conference on Applied Machine Learning in Information Security (CAMLIS), October 19--20, 2023, Arlington, VA}

\title{MalDICT: Benchmark Datasets on Malware Behaviors, Platforms, Exploitation, and Packers}

\author[1,2,3]{Robert J. Joyce}[%
email=joyce8@umbc.edu
]
\author[1,2]{Edward Raff}[%
email=raff.edward@gmail.com
]
\author[3]{Charles Nicholas}[%
email=nicholas@umbc.edu
]
\author[1]{James Holt}[%
email=holt@lps.umd.edu
]

\address[1]{Laboratory for Physical Sciences}
\address[2]{Booz Allen Hamilton}
\address[3]{University of Maryland Baltimore County}

\newif\iflong\longfalse
\newcommand\longv[1]{\iflong {\color{blue} #1}\else\fi}

\begin{abstract}
Existing research on malware classification focuses almost exclusively on two tasks: distinguishing between malicious and benign files and classifying malware by family. However, malware can be categorized according to many other types of attributes, and the ability to identify these attributes in newly-emerging malware using machine learning could provide significant value to analysts. In particular, we have identified four tasks which are under-represented in prior work: classification by behaviors that malware exhibit, platforms that malware run on, vulnerabilities that malware exploit, and packers that malware are packed with. To obtain labels for training and evaluating ML classifiers on these tasks, we created an antivirus (AV) tagging tool called ClarAVy. ClarAVy's sophisticated AV label parser distinguishes itself from prior AV-based taggers, with the ability to accurately parse 882 different AV label formats used by 90 different AV products. We are releasing benchmark datasets for each of these four classification tasks, tagged using ClarAVy and comprising nearly 5.5 million malicious files in total. Our malware behavior dataset includes 75 distinct tags - nearly 7$\times$ more than the only prior benchmark dataset with behavioral tags. To our knowledge, we are the first to release datasets with malware platform and packer tags. 

\end{abstract}


\maketitle

\section{Introduction}
\label{sec:introduction}

The malware ecosystem is both massive and diverse. Novel malware emerges regularly and existing malware is continually being updated to add functionality or improve evasion \cite{talukder2020survey}. Analyzing malware by hand is slow and requires expert domain knowledge, so machine learning and other forms of automation are relied upon as a supplement \cite{mohaisen2013,10.1145/3290607.3313040}. As a result, there has been significant research effort towards improving malware classification using machine learning. Existing work almost exclusively focuses on two classification problems: malware detection (detecting whether a file is malicious or benign) and malware family classification (determining the malware family that a malicious file belongs to) \cite{raff2020survey}. To our knowledge, SOREL is the only  malware benchmark dataset that is currently available to the public and provides labeled data for a different classification problem than the two listed above. SOREL contains $\approx$10 million malicious files but is labeled according to just 11 behavioral tags \cite{sorel}. In the wild, malware exhibits a far greater variety of behaviors, and there are other attributes by which malware can be classified that are entirely unexplored. The Malicia dataset includes 11,363 malware samples tagged according to 172 distinct exploits, but it is nine years old as of the time of writing and is no longer publicly distributed \cite{malicia}.

We have assembled a collection of four benchmark datasets named MalDICT (Malware Datasets for Infrequent Classification Tasks), with each dataset supporting a different, under-represented malware classification task. MalDICT is being published in the hope that it will encourage increased awareness and study of these tasks. We are also publishing benchmark results after training two standard malware classifiers on each of these datasets. This will enable researchers to compare the performance of their own models against a baseline and against each other.
The four benchmark datasets within MalDICT are:

\begin{enumerate}
    \item \textbf{MalDICT-Behavior:} 4,317,241 files tagged according 75 common malware categories and malicious behaviors. 

    \item \textbf{MalDICT-Platform:} 963,492 files tagged according to 43 common file formats, operating systems, and programming languages.

    \item \textbf{MalDICT-Vulnerability:} 173,886 files tagged according to 128 common vulnerabilities exploited by malware.

    \item \textbf{MalDICT-Packer:} 252,148 files tagged according to 79 common malware packers.
\end{enumerate}

\begin{wrapfigure}[17]{R}{0.55\textwidth}
\vspace*{-28pt}
    \centering
    \captionsetup{format=plain}
    \includegraphics[width=0.45\columnwidth,keepaspectratio]{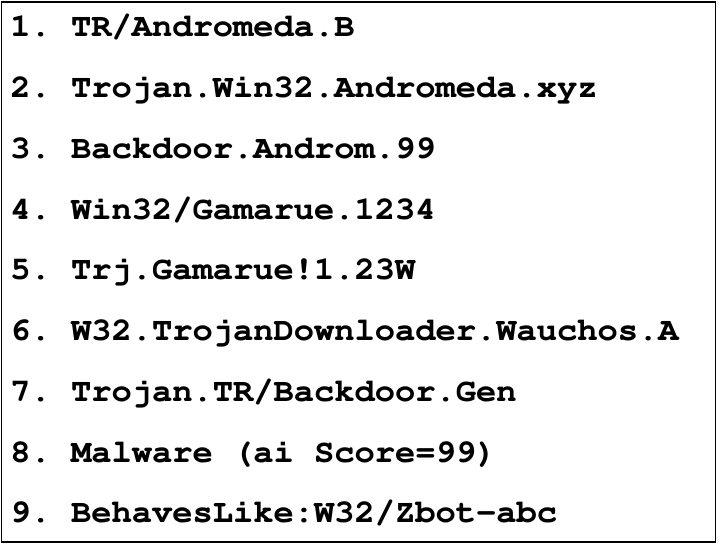}
    \caption{Fictitious AV scan report for a file. Labels 1-6 correctly classify this file as belonging to the Andromeda family, which has the aliases ``Androm", ``Gamarue", and ``Wauchos". Label 7 does not assign the malware a family, but indicates that the it belongs to the Trojan and Backdoor categories. Labels 8-9 are heuristics, and Label 9 incorrectly classifies the malware into the Zeus family.}
    \label{fig:scan-results}
    
\end{wrapfigure}

\subsection{Antivirus Terminology}

MalDICT was tagged by combining outputs from multiple different antivirus (AV) products. We developed a custom tool named ClarAVy for this, which we describe in Section \ref{sec:claravy}. For the remainder of this section, we introduce terminology about AV products and survey related AV-based taggers. 

When detecting a file as malware, an AV product will produce an output called an \textbf{\textsl{AV label}}. An example AV label is \texttt{Trojan:}\texttt{Win32.}\texttt{Androm}\texttt{.abc}. Each portion of the label describes a characteristic of the file that the AV detected as malicious, such as its behavior, file format, or family. In some cases AV labels may also include a threat group the malware is attributed to, a vulnerability the malware exploits, or the packer the file was packed with \cite{avclass2}. Scanning a malicious file with a collection of AV products generates an \textbf{\textsl{AV scan report}}. An example AV scan report is shown in Figure \ref{fig:scan-results}. Note that the naming conventions and label formats used by each AV product are different. Also note the tokens with different spellings but identical meanings, such as W32/Win32, TR/Trj/Trojan, and Andromeda/Androm. We call these \textbf{\textsl{token aliases}}. 

\subsection{Related Work}
\label{sec:related-work}

AVClass is the seminal work on labeling malware using AV scan reports \cite{avclass}. Given a report, the tool filters out duplicate AV labels, normalizes and tokenizes each label, filters out non-family tokens, and renames families that have known aliases. The most common remaining token becomes the family tag for the scan report. 
AVClass++ \cite{avclassplusplus}, Sumav \cite{sumav}, and AVMiner \cite{avminer} are other tools which output malware family tags using AV scan data.

To our knowledge, the following AV-based taggers can assign non-family tags to malware. 
EUPHONY creates a graph with weighted edges between related reports, forms clusters from communities in the graph, and assigns labels based on the majority family, category, or file type in the cluster \cite{euphony}.
SMART distills AV scan reports into a multi-label representing the file's behaviors \cite{SMART}. However, it supports only 11 malicious behaviors. It is the tagging method used by the SOREL dataset. AVClass2  is an update to AVClass by its original creators \cite{avclass2, avclass}. It identifies tokens in AV scan reports which indicate the family, category, file type, and notable behaviors of the malware. It is the only tool we surveyed which can identify tokens related to packers and  vulnerabilities, but it relies on a hard-coded list of tokens to do so. Finally, Garc{\'\i}a-Teodoro et al. \cite{multilabeling} created a tool that outputs multi-labels corresponding to the counts of behavioral tokens in AV scan reports.

\section{ClarAVy}
\label{sec:claravy}

The purpose of ClarAVy is to clarify the

\begin{wrapfigure}[7]{R}{0.55\textwidth}
\vspace*{-56pt}
    \centering
\includegraphics[width=0.52\columnwidth,keepaspectratio]{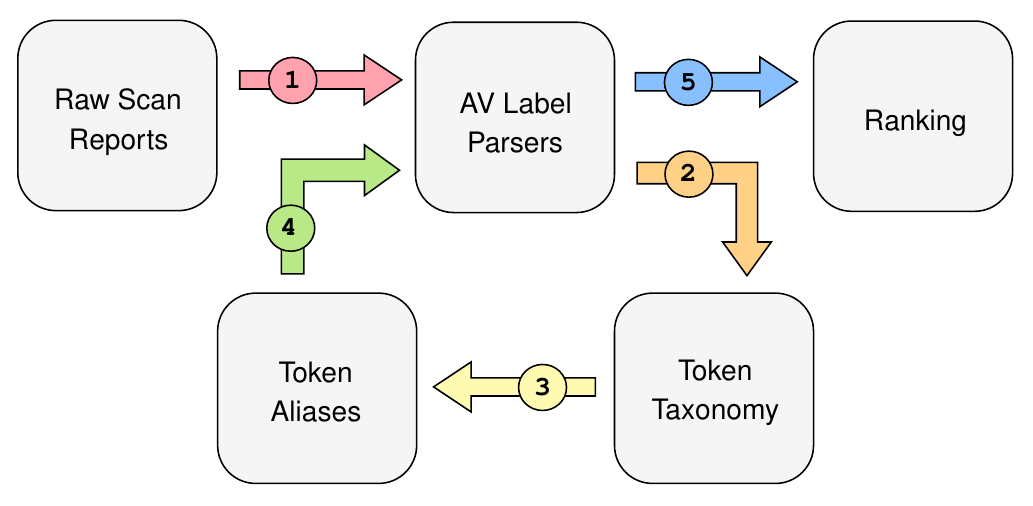}
    \caption{ClarAVy Architecture.}
    \label{fig:claravy-architecture}
\end{wrapfigure}

\noindent  noisy outputs from a collection of AV products into simple, easy-to-interpret tags. It is the tool used for labeling the malware in MalDICT. Figure \ref{fig:claravy-architecture} shows the major stages of the ClarAVy architecture. First, ClarAVy ingests a corpus of AV scan reports and tokenizes each label. A lexical category is assigned to most tokens (i.e., whether the token indicates a malware  behavior, packer, etc.). Once all scan reports are processed, ClarAVy reviews any tokens with incomplete or ambiguous parsing and attempts to assign a global lexical category to them. Next, ClarAVy identifies tokens which are aliases of each other. Finally, ClarAVy re-processes all of the scan reports, this time using its newly-obtained information about lexical assignments and token aliases. For each scan report, ClarAVy outputs a token ranking and the lexical category each token was assigned to. In the remainder of this section, we provide technical details for each of these stages.

\subsection{Token Taxonomy}

A taxonomy of the different lexical categories which ClarAVy can assign to tokens is provided in Table \ref{tab:taxonomy}. The FAM, PLAT, and BEH lexical categories are analogous to the "Family", "Platform", and "Type" fields in the CARO malware 

\begin{wraptable}[10]{R}{0.55\textwidth}
\centering
\vspace*{-24pt}
\caption{Taxonomy of Tokens in AV Labels}
\label{tab:taxonomy}
\adjustbox{max width=0.52\textwidth}{

\begin{tabular}{@{}l@{}l@{}}
\toprule
BEH & The malware category or behavior \\
\midrule
PLAT & The OS, file format, or programming language \\
\midrule
VULN & A vulnerability exploited by the malware\\
\midrule
PACK & The packer used to pack the file \\
\midrule
FAM & The malware family that the file belongs to \\
\midrule
SUF & A suffix token at the end of the AV label \\
 \midrule
PRE & Ambiguous, but not a FAM or SUF token \\
 \midrule
UNK & A token whose lexical category cannot be determined \\

\bottomrule
\end{tabular}
}
\vspace*{6pt}

\end{wraptable}

\noindent naming scheme. Some AV labels include additional information, such as the packer used to pack the file (PACK) or a vulnerability the malware exploits (VULN). Some AV labels may not be able to be fully parsed, necessitating the PRE and UNK lexical categories. The suffix of the AV label is assigned the SUF lexical category.

\subsection{AV Label Parsing}

While parsing AV scan reports, ClarAVy

\begin{wrapfigure}[21]{R}{0.55\textwidth}
\vspace*{-38pt}
    \centering
    \captionsetup{format=plain}
\includegraphics[width=0.52\columnwidth,keepaspectratio]{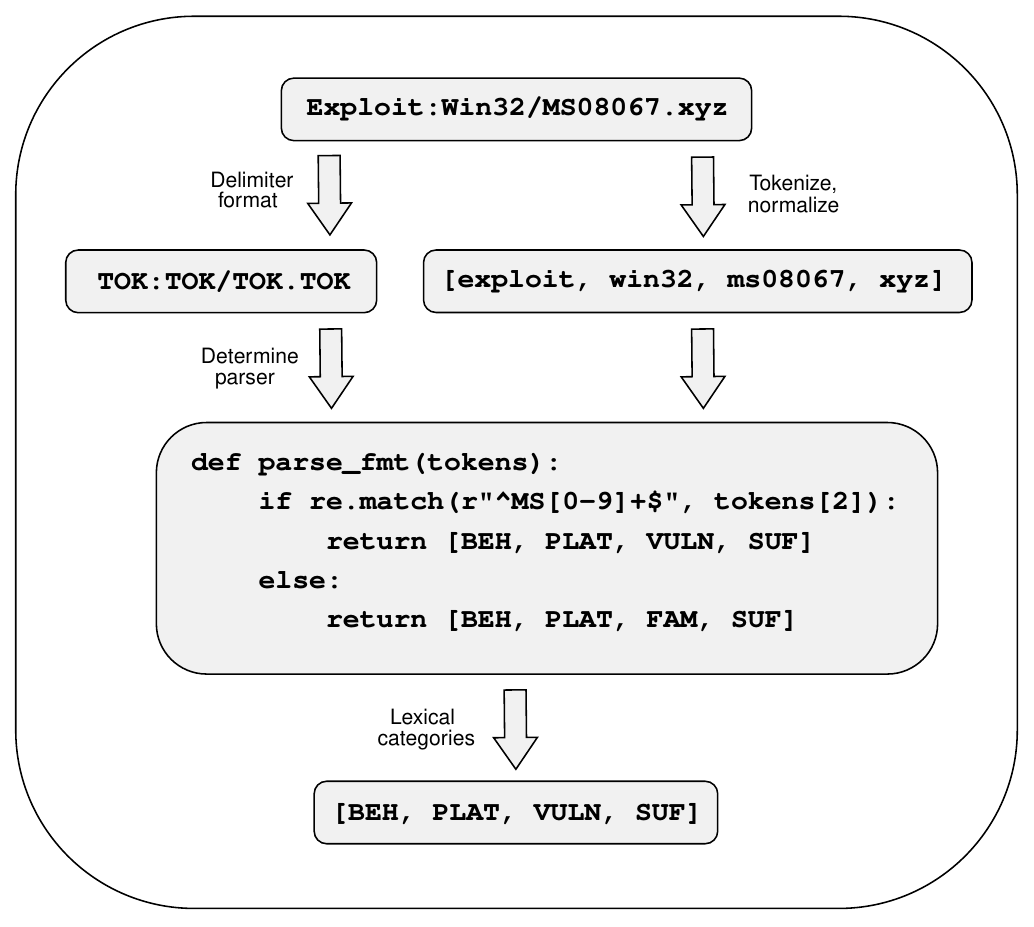}
    \caption{Parsing of the label ``Exploit:Win32/MS08067.xyz". ClarAVy identifies the delimiter format of the label and selects a parsing function for it. This simple parsing function distinguishes between AV labels in this delimiter format that contain VULN or FAM tokens. The assigned lexical categories indicate that this AV label detects exploitation of the MS08-067 vulnerability in Windows.}
    \label{fig:parsing}
\end{wrapfigure}

\noindent identifies tokens within AV labels and attempts to assign each token to a lexical category in its taxonomy. ClarAVy tokenizes AV labels by splitting them on "delimiter" tokens, which are any tokens in the label that are non-alphanumeric. We call the sequence of delimiter tokens in an AV label its \textbf{delimiter format}. Most AV labels with the same delimiter format and from the same AV product have lexical categories in predictable locations. ClarAVy takes advantage of this property to make parsing simpler while also reducing ambiguity. 

After identifying the delimiter format for an AV label, ClarAVy selects an appropriate parsing function. The parsing function attempts to assign a lexical category to each token in the label. Most of ClarAVy's parsing functions do this by applying  regular expressions and boolean logic to the tokens. Figure \ref{fig:parsing} shows how the AV label \texttt{Exploit:Win32/MS08067.xyz} is parsed. This label is applied to malware which exploits the MS08-067 vulnerability. The delimiter format for this label is \texttt{TOK:TOK/TOK.TOK}, where \texttt{TOK} represents the locations that tokens may appear in the label. AV labels that have this delimiter format always have a CAT token in the first position, a TGT token in the second position, and a SUF token in the fourth position. A token in the third position of the label may either be a VULN or a FAM token, and the parsing function uses a regular expression to determine which lexical category should be assigned to it.

ClarAVy includes parsing functions for \textbf{882 different delimiter formats} across \textbf{90 AV products} -  over 8,000 lines of Python code in total. A few AV products use only a single delimiter format, while others have dozens. Parsing functions range from trivial to complex, depending on how standardized the labels of an AV product are. We identified the most common delimiter formats used by each AV product to ensure maximal coverage. Then, we manually implemented and verified each parsing function to ensure that the lexical categories ClarAVy assigns to tokens are accurate.

\subsubsection{Handling Parsing Ambiguity}

In some cases, ClarAVy's parsing functions cannot assign lexical categories to some tokens in an AV label. This is most often due to there being no programmatic way to distinguish tokens indicating behavior, platform, vulnerability, and/or packer from each other or from other generic tokens. The parsing function assigns these tokens the PRE lexical category to indicate that there is some ambiguity, but it is not a FAM or SUF token. More rarely, there are edge cases where tokens are truly ambiguous. The parsing function assigns the UNK lexical category to these tokens.

After all scan reports are parsed for the first time, ClarAVy attempts to determine the lexical category of each token that had some parsing ambiguity. Even if a token is assigned PRE or UNK by one parsing function, it may appear in other AV labels where it can be parsed correctly. If a token is unanimously assigned to a lexical category (not counting PRE and UNK), it is permanently assigned to that category when ClarAVy is used in the future. 

ClarAVy is provided with a default wordlist that maps tokens to their lexical categories. This wordlist was generated by running ClarAVy on $\approx$40 million AV scan reports from VirusTotal \cite{virustotal}. We describe how we collected these scan reports in Section \ref{sec:validation}. Users can add to or alter this wordlist if they have different preferences. For example, we manually removed the ``trojan" and ``win32" tokens from the BEH and PLAT categories, respectively, since they are nearly ubiquitous in AV scan reports. In particular, AV products tend to use the ``trojan" tag generically rather than for actual trojan malware \cite{avclass2}.

\subsection{Token Alias Resolution}

During its next stage, ClarAVy attempts to identify tokens that have identical meaning. In FAM tokens, aliases may have very distinct spellings (e.g. Andromeda, Gamaue, and Wauchos in Figure \ref{fig:scan-results}). However, we observe that aliases for tokens in most other lexical categories generally have similar spellings. Our approach to token alias resolution uses a metric based on edit distance in addition to token co-occurrence percentage. We identify two different classes of token aliases, which we call \textbf{trivial aliases} and \textbf{parent-child aliases}.

\subsubsection{Identifying Trivial Aliases}
We say that a pair of tokens are trivial aliases if they share a lexical category and are nearly identical in spelling, where a single minor edit can transform one token into the other. For example, if one token can be transformed into a second token adding extra digit or character to the end (e.g. ``backdoor" and ``backdoor0"), ClarAVy considers the pair to be trivial aliases. Additionally, ClarAVy uses a small list of common substrings that are frequently observed at the beginning and end of tokens. If two tokens are identical except for the substring, it assigns them as aliases. Trivial aliases are very frequent in AV scan data, and this procedure is simple but highly effective.

\subsubsection{Identifying Parent-Child Alias Candidates}
ClarAVy also recognizes aliases from token pairs which have a ``parent-child" relationship. Two conditions must apply to satisfy this relationship. First, the less common token (the child) must co-occur with the more common token (the parent) in a sufficient percentage of scan reports. Additionally, a score based on edit distance must be sufficiently high. We adapt metrics from Sebasti{\'a}n et al. \cite{avclass2} for computing co-occurrence percentage between tokens. Let the number of scan reports containing the child token be given by $|t_{i}|$, and let $|(t_{i}, t_{j})|$ be the number of scan reports containing both the child and parent token. The frequency that the child token co-occurs with the parent token is given by:

\begin{equation*}
\coocur(t_{i}, t_{j}) = \frac{|(t_{i}, t_{j})|}{|t_{i}|}
\end{equation*}\

A high co-occurrence percentage indicates that the child token may be related to the parent token, but other factors (such as spurious correlations between the outputs of different AV products) may cause dissimilar tokens to co-occur frequently. To reduce false positives, we also require that pair of tokens is similar in spelling. Let $\len(t)$ be the number of characters in token $t$. We define a custom edit score based on edit distance:

\begin{equation*}
    \escore(t_{i}, t_{j}) = 1 - \edist(t_{i}, t_{j}) \; /  \; \min(\len(t_{i}), \len(t_{j}))
\end{equation*}\

Afterwards, we apply two heuristics to the edit score which we frequently observe in token aliases. If the shorter token is a substring in the longer token, or if the two tokens are anagrams, the edit score is capped at a minimum of 0.75. ClarAVy uses threshold parameters $E$ (0.6 by default) and $C$ (0.5 by default) to control parent-child aliasing. If $\escore(t_{i}, t_{j}) >= E$ and $\coocur(t_{i}, t_{j}) \times \escore(t_{i}, t_{j}) >= C$, then $t_{i}$ has a parent-child relationship with $t_{j}$.

\subsubsection{Resolving Parent-Child Aliases}

Pairs of tokens with parent-child relationships are not immediately considered to be aliases. This is because a token may share a parent or child relationship with multiple other tokens. Algorithm \ref{alg:parent-child} shows how ClarAVy identifies aliases from the set of tokens with parent-child relationships. Let $T$ be a list of all known tokens within the same lexical category (e.g. all of the BEH tokens) sorted by token frequency, descending. At each iteration of the algorithm, the cur-

\begin{wrapfigure}[20]{R}{0.5\textwidth}
\vspace{-12pt}
\begin{minipage}{0.47\textwidth}
\begin{algorithm}[H]
\caption{Parent-Child Alias Resolution}
\label{alg:parent-child}
\begin{algorithmic}[1]
\Require Sorted list of tokens $T$

\Function{Alias\_Resolve}{$T$} 
    \State $U \leftarrow \emptyset$
    \For {$t_{i} \in T$}
        \State $A \leftarrow \emptyset$
        \State $Q \leftarrow \queue$
        \State $Q.\enqueue(t_{i})$
        \While {not $Q.\isempty()$}
            \State $t_{j} \leftarrow Q.\dequeue()$   
            \If {$t_{j} \notin U$ and $t_{j} \notin A$}
                \State $A \leftarrow A \cup t_{j}$
                \For {$t_{k} \in t_{j}.\children$}
                    \State $Q.\enqueue(t_{k})$
                \EndFor
            \EndIf
        \EndWhile
        \For {$t_{j} \in A$}
            \State $U \leftarrow U \cup t_{j}$
            \If {$t_{i} \neq t_{j}$}
                \State $\resolvepair(t_{i}, t_{j})$
            \EndIf 
        \EndFor
    \EndFor    
\EndFunction

\end{algorithmic}
\end{algorithm}
\captionsetup{format=plain}
\caption{Algorithm for selecting aliases from parent-child candidates.}
\end{minipage}
\end{wrapfigure}

\noindent rent token $t_i$ is treated as the \textbf{canonical name} for all of its aliases (i.e., all of its aliases will be  renamed to the current token). A set of all ``descendants" of the current token is created by recursively visiting child tokens. Then, each descendent token is assigned as an alias of the current token, provided that it has not been assigned a different alias already.

By default, the canonical name for a group of aliases is the most frequently-appearing token in the provided AV scan report dataset. ClarAVy comes with a text file which maps tokens to their canonical alias names. It was generated by using the previously-described alias resolution process on a dataset of $\approx$40 million AV scan reports from VirusTotal \cite{virustotal}. ClarAVy users can easily customize this mapping with their own alias pairs by editing the file. Furthermore, the canonical names in the alias mapping have priority over automatically-identified canonical names, allowing users to 

\noindent set their naming preferences.

\subsection{Token Ranking}

After assigning lexical categories to tokens and after resolving aliases, ClarAVy parses all scan reports a second time. This time, tokens with known aliases are replaced with their canonical names. Additionally, tokens which were previously assigned PRE or UNK may receive a more informative lexical category. For each AV  scan report, ClarAVy outputs a ranking the BEH, PLAT, PACK, and VULN tokens in the report. 
Each token is given a score based on the number of times it appears in the scan report, adjusted for known correlations between AV products. PRE, SUF, and UNK tokens in the scan report are considered generic and discarded. Accurately ranking 
 FAM tokens is much more challenging and is a target of our future work.

\subsubsection{AV Product Correlations}
The existence of correlations between AV products is well-known in the malware analysis industry. Leading causes include AV products sub-licensing their engines to others, AV products owned by the same company, and AV products ``copying" another product's detection results \cite{avclass, av-meter}. There seem to be other factors contributing to these correlations as well, but they are poorly-understood \cite{joyce2021rank1}. We attempted to account for all major, publicly-known factors which would cause AV products in our dataset to produce correlated labels. To do this, we identified AV products which use very similar sets of delimiter formats in their labels. We manually confirmed each pair of correlated AV products that we wrote parsing rules for using publicly-

\begin{wrapfigure}[17]{R}{0.55\textwidth}
    \centering
\includegraphics[width=0.53\columnwidth,keepaspectratio]{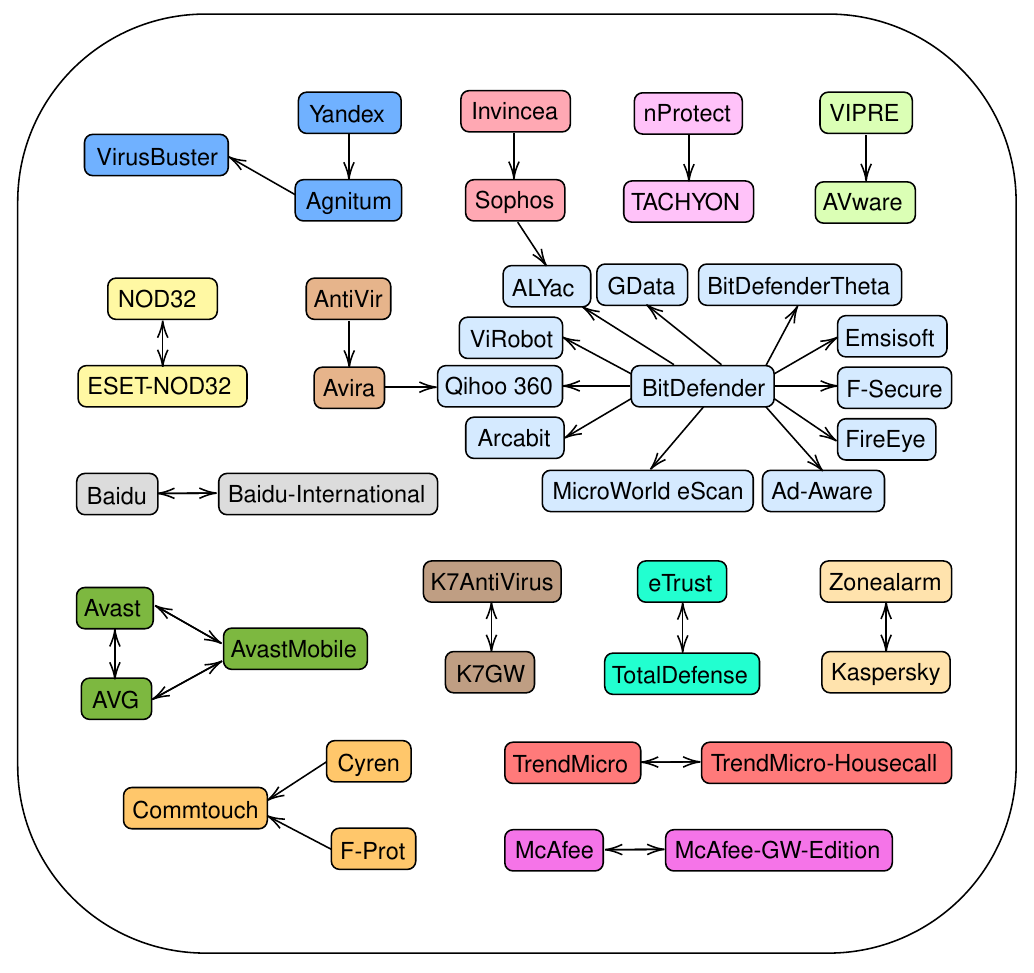}
    \caption{Publicly-known correlations in AV products.}
    \label{fig:av-correlation}
\end{wrapfigure}

\noindent available information, shown in Figure \ref{fig:av-correlation}. Like prior work, we observed that the main sources of correlation were due to AV products owned by the same company (e.g. McAfee and McAfee-GW-Edition) and AV products licensing their technology to others (e.g. ZoneAlarm previously used Kaspersky's engine). In our dataset, there are 11 different AV products which use the BitDefender engine to varying degrees, often in combination with their own detection technologies. ALYac and Qihoo 360 use multiple other engines. We also noticed multiple instances of AV products being renamed or acquired by other companies (e.g. Commtouch was renamed to Cyren and aquired F-Prot).

\begin{wrapfigure}[15]{R}{0.55\textwidth}
\vspace*{-36pt}
    \centering
    \captionsetup{format=plain}
\includegraphics[width=0.53\columnwidth,keepaspectratio]{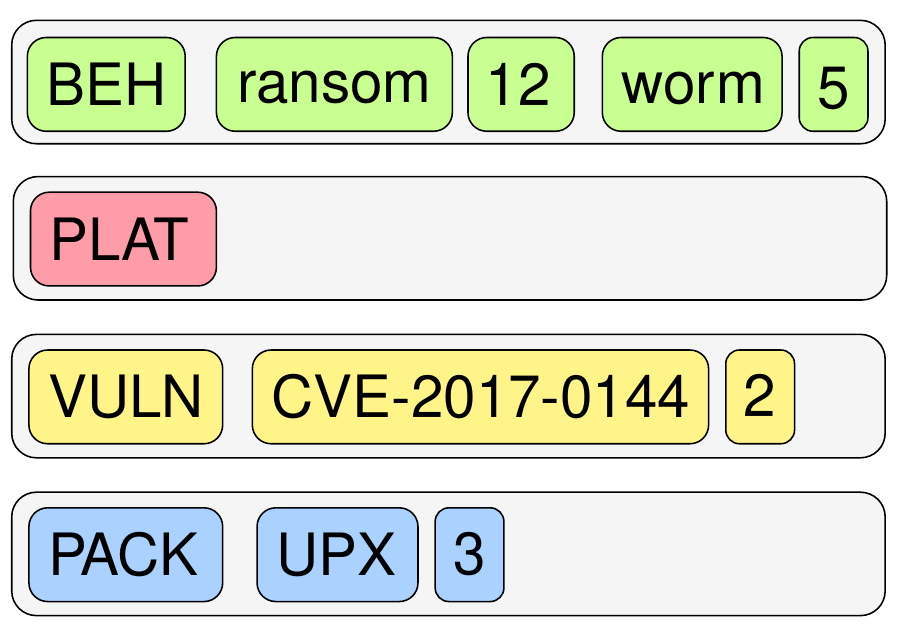}
    \caption{Example ClarAVy output for a malicious file. It is tagged as having ransomware and worm behavior. The file exploits the CVE-2017-0144 vulnerability and is packed with UPX. No platform tags are identified.}
    \label{fig:parsing}
\end{wrapfigure}

\subsubsection{Token Scores}

ClarAVy assigns a   score to each token based on the number of times it appears in the scan report. This approach assumes that if multiple independent AV products output the same token, then the token is likely to be an accurate tag for the file. For this assumption to be valid, correlations between AV products must be accounted for. If two or more AV products with known correlations output the same token, ClarAVy combines them into a single ``vote". 

A threshold parameter $T$ is used to control the minimum token score allowed in the ranking that ClarAVy outputs. Tokens with fewer than $T$ votes are excluded from the ranking. Higher values of $T$ decrease the amount of noise in the outputs, but may also cause correct tokens to be omitted from the ranking. $T$ can be set separately for each lexical category. By default $T=5$ for BEH and PLAT tokens and $T=1$ for VULN and PACK tokens. As we later show in Section \ref{sec:sorel-val}, agreement of at least 5 independent AV products has a very low false positive rate. VULN and PACK tokens are much less frequent than BEH and PLAT tokens, but also much less noisy. Any threshold above $T=1$ for these lexical categories would cause a high false negative rate.

\begin{wraptable}[11]{R}{0.55\textwidth}
\centering
\vspace*{-44pt}
\caption{Top 10 Tokens Per Lexical Category}
\label{tab:token-stats}
\adjustbox{max width=0.52\textwidth}{
\begin{tabular}{@{}llll@{}}
\toprule
 BEH & PLAT & VULN & PACK \\
\midrule
virus & js & cve\_2014\_6332 & nsis \\
downloader & script & cve\_2010\_2568 & upx  \\
riskware & html & cve\_2017\_17215 & nsanti    \\
adware & pe & cve\_2017\_11882 & upack  \\
dropper & vbs & cve\_2010\_0188 & aspack  \\
pua & hllo & cve\_2017\_0199 & themida    \\
packed & msil & cve\_2010\_2586  & nspack   \\
worm & pdf & cve\_2010\_2586  & pecompact  \\
backdoor & multi & cve\_2012\_0507  & fsg \\
redirector & android & cve\_2012\_0507  & vmprotect \\
\bottomrule
\end{tabular}
}
\vspace*{6pt}
\end{wraptable}

\section{ClarAVy Validation}
\label{sec:validation}
ClarAVy was developed with AV scan reports for 40,307,433 malicious files from chunks 0 through 465 of the VirusShare corpus \cite{virusshare}. We queried the VirusTotal API for these files between Feb. and Apr. 2023 to get these reports \cite{virustotal}. When developing each parsing function in ClarAVy, we randomly selected 10,000 AV labels with the corresponding delimiter format from this dataset. After creating a parsing function, we performed a brief visual inspection of the resulting tokens and lexical assignments to ensure they were correct. After finishing the entire ClarAVy implementation, we ran it on these $\approx$40 million AV scan reports with default settings ($T=5$ for BEH and PLAT tokens and $T=1$ for VULN and PACK tokens). Then, we inspected lexical categories that ClarAVy assigned to each token and the alias mapping which it created. We manually verified both of these, correcting any errors if necessary. We identified 1,307 aliases for 92 malware behaviors, 194 aliases for 47 file-related tokens, and 53 aliases for 24 packers. The ClarAVy output included 134 distinct BEH tokens, 91 distinct PLAT tokens, 440 distinct VULN tokens, and 90 distinct PACK tokens. The ten most common tokens of each type are listed in Table \ref{tab:token-stats}.

\subsection{Comparison to other AV-based taggers}

ClarAVy's comprehensive collection of parsing functions distinguishes it from other AV-based taggers. Most prior work uses hard-coded lists and/or heuristic methods for assigning tokens to lexical categories \cite{avclass, avclassplusplus, avclass2, euphony}. For example, AVClass2 uses one parsing function per AV product for removing the suffix from AV labels \cite{avclass2}. Then, it uses hard-coded lists for assigning remaining tokens in the to lexical categories. AVClass2 supports updating these lists with related tokens, but it is not a default behavior and uses only co-occurrence statistics. These design choices lead to compounding errors in AVClass2's outputs. Using the same method for suffix removal on all of an AV product's labels may cause incorrect parsing, since the AV product likely has multiple delimiter formats. Using only hard-coded lists for assigning lexical categories will result in false negatives - especially if new tokens appear in future AV labels. With 882 parsing functions (averaging nearly 10 per supported AV product), ClarAVy assigns lexical categories to AV labels with greater fidelity. It can handle new AV labels, provided that their delimiter formats are supported.

\subsection{Evaluation Using the SOREL Dataset}
\label{sec:sorel-val}

We experimentally test ClarAVy's ability to tag malware according to behavioral attributes. We do this using the SOREL dataset, which has 9,919,065 malicious PE files labeled according to 11

\begin{wraptable}[7]{R}{0.55\textwidth}
\centering
\vspace*{-5pt}
\caption{SOREL Evaluation (Micro Avg.)}
\label{tab:sorel-micro}
\adjustbox{max width=0.52\textwidth}{
\begin{tabular}{@{}lrrr@{}}
\toprule
 & ClarAVy (T=1) & ClarAVy (T=5) & AVClass2 \\
\midrule
Precision & .663 & \textbf{.969} & .785\\
Recall & \textbf{.625} & .251 & .483\\
F1-Measure & \textbf{.643} & .398 & .598 \\
\bottomrule
\end{tabular}
}
\vspace*{6pt}
\end{wraptable}

\noindent  separate behavioral tags \cite{sorel}. A file may have more than one tag if it displays multiple types of malicious behaviors. We queried the VirusTotal API for the malicious files in SOREL and were able to obtain AV scan reports for 7,294,655 of them. Then, we ran ClarAVy on these re-

\begin{wraptable}[7]{R}{0.55\textwidth}
\centering
\vspace*{-8pt}
\caption{SOREL Evaluation (Weighted Avg.)}
\label{tab:sorel-weighted}
\adjustbox{max width=0.52\textwidth}{
\begin{tabular}{@{}lrrr@{}}
\toprule
 & ClarAVy (T=1) & ClarAVy (T=5) & AVClass2 \\
\midrule
Precision & .717 & \textbf{.970} & .830\\
Recall & \textbf{.625} & .251 & .483\\
F1-Measure & \textbf{.668} & .398 & .610 \\
\bottomrule
\end{tabular}
}
\vspace*{6pt}
\end{wraptable}

\noindent  ports two times; once with $T=1$ for all lexical categories and once with $T=5$ for BEH and PLAT tokens. We also ran AVClass2 on these reports using default settings, except for an adjustment to its alias mapping which removes the alias between the ``dropper" and ``downloader" tokens. This is because SOREL treats these as seperate tags, but AVClass2 does not by default. It was also necessary to adjust the naming for some tags, since ClarAVy, AVClass2, and SOREL use slightly different terminology. We measured the per-class Precision, Recall, and F1-Measure for each of the 11 behavioral tags. Results are shown in Table \ref{tab:sorel-micro} (with micro averaging) and Table \ref{tab:sorel-weighted} (with weighted averaging). ClarAVy with $T=1$ achieves the highest Recall and F1-measure, but has the lowest precision. AVClass effectively uses $T=2$, since it discards any tokens which only recieve a single vote. This allows it to reach a higher Precision than ClarAVy with $T=1$, but the Recall and F1-Measure drop because some correct labels are discarded. ClarAVy with $T=5$ reaches an extremely high Precision but a low Recall and F1-Measure for the same reason. The very low false positive rate of ClarAVy with $T=5$ is a desirable property for an accurately-tagged dataset, and we judge the false negative rate to be of little impact.

\subsection{Evaluation Using the MOTIF Dataset}
\label{sec:motif-val}

Labeled malware data which can be used to evaluate ClarAVy is extremely limited. With the exception of SOREL, nearly all malware reference datasets either use benign/malicious labels or family labels \cite{raff2020survey}. SOREL only has 11 behavioral tags, does not have labels comparable to the PLAT, VULN, or PACK lexical categories that ClarAVy uses, and is itself dependent on AV scan data (due to using SMART as a source of labeling) \cite{sorel, SMART}. Therefore, it was necessary to find another way to evaluate ClarAVy's outputs. To do this, we consider that malicious files belonging to the same family should be consistent regarding malware category, behavior, file format, and other factors. Although this is not always true (e.g., modular malware in the same family may have different components with specialized behavior, and in rare cases malware is written to target different platforms), this assumption generally holds. This allows us to evaluate how consistent ClarAVy's outputs are with respect to malware family labels.

Suppose a dataset of malicious files $M = \{m_1, m_2, ...m_n$\}, where $n$ is the number of files in the dataset. Let $C_{i} \in M$ be the set of files that a malware tagging tool assigns tag $i$. For example, $C_{i}$ could be the set of files that ClarAVy assigns the ``ransomware" tag to. A malicious file may be assigned multiple tags. Then, let $F = \{F_{k}\}_{1\leq k \leq f}$ partition $M$, where $F_k$ is the set of malicious files belonging to family $k$. Each file is assigned to exactly one family. Then, for each predicted label $C_i$, let $D_{i} = \bigcup\limits_{k=1}^{f}  F_k, \; \textrm{if} \;\; \frac{|C_i \cap F_k|}{|F_k|} \geq 0.5 $. This constructs a set of malicious files $D_{i}$ from malware families where at least 50\% of files have tag $i$ predicted. Using this, we can define metrics which are analogous to per-tag Precision and Recall:

\[ Precision(C, D, i) =  \frac{|C_i \cap D_i|}{|C_i \cap D_i| + |C_i \cap \bar{D_i}|}\;\;\;\;\;\;\;\;\;\;\;\;\;\; Recall(C, D, i) =  \frac{|C_i \cap D_i|}{|C_i \cap D_i| + |\bar{C_i} \cap D_i|} \]

\vspace*{12pt}

Under these definitions, Precision   measures the ``noisiness" of a tag. It penalizes instances where a tag is assigned to a file, but where most files in its family are not associated with that tag. Conversely, Recall measures coverage of a tag within malware families. It penalizes instances where a family is likely to be associated with that tag, but there are files within that family where the tag is not assigned.

\begin{wraptable}[7]{R}{0.50\textwidth}
\centering
\vspace*{-16pt}
\caption{MOTIF Evaluation (Micro Avg.)}
\label{tab:motif-micro}
\adjustbox{max width=0.47\textwidth}{
\begin{tabular}{@{}lrr@{}}
\toprule
& ClarAVy & AVClass2 \\
\midrule
Precision & \textbf{.828} & .694\\
Recall & \textbf{.912} & .796 \\
F1-Measure & \textbf{.868} & .741 \\
\bottomrule
\end{tabular}
}
\vspace*{6pt}
\end{wraptable}

We acknowledge that there are flaws in this evaluation strategy. It is possible that families which are truly associated with a tag may be ``missed" due to incorrect predictions. Furthermore, it does not necessarily confirm that predicted tags are correct (although we believe this is likely in most instances due to the high precision observed in our previous exp-

\begin{wraptable}[7]{R}{0.50\textwidth}
\centering
\vspace*{-24pt}
\caption{MOTIF Evaluation (Weighted Avg.)}
\label{tab:motif-weighted}
\adjustbox{max width=0.47\textwidth}{
\begin{tabular}{@{}lrr@{}}
\toprule
& ClarAVy & AVClass2 \\
\midrule
Precision & \textbf{.880} & .723\\
Recall & \textbf{.912} & .796 \\
F1-Measure & \textbf{.896} & .758 \\
\bottomrule
\end{tabular}
}
\vspace*{6pt}
\end{wraptable}

\noindent eriment using the SOREL dataset). However, in the absence of better-labeled data, we believe that this is a reasonable approach for measuring ClarAVy's tagging consistency.

We then used ClarAVy and AVClass2 to tag VirusTotal reports for the MOTIF dataset. MOTIF contains 3,095 malware samples from 454 families, labeled with ground-truth confidence. ClarAVy and AVClass2 were run on default settings, and any tags in AVClass2's family (FAM) or unknown (UNK) taxonomy were discarded because they are not output by ClarAVy. Additionally, the AVClass2 ``windows" tag was discarded, since it appears in nearly all scan reports in MOTIF and ClarAVy treats it as generic. We computed Precision, Recall, and  F1-Measure for each label using the method described above. Results are show in Tables \ref{tab:motif-micro} and \ref{tab:motif-weighted}. ClarAVy clearly outperforms AVClass2 in this experiment. ClarAVy tags malware within the same family more consistently and there is less tagging noise.

\section{MalDICT Datasets}
\label{sec:datasets}

To build the MalDICT datasets, we ran ClarAVy on 40,307,433 VirusTotal reports for the malware in VirusShare chunks 0-465. We reviewed the tags that ClarAVy had assigned to these files and observed significant class imbalances. To account for this, we discarded tags which were too rare and down-sampled tags which were very common. BEH tags with less than 1,000 instances, PLAT tags with less than 500 instances, VULN tags with less than 100 instances, and PACK tags with less than 50 instances were not included. Tags which were too frequent were randomly down-sampled, so that they were no more than 100$\times$ more common than the minimum threhsold in the training set and no more than 25$\times$ more common than the minimum threshold in the test set.

Depending on the lexical category, we selected between two different methods for dividing files into a training and test set. For MalDICT-Behavior and MalDICT-Platform, we selected from files in VirusShare chunks 0-315 for the training set and from VirusShare chunks 316-465 for the test set. Chunks 0-148 contain 131,072 files each, and the remaining chunks contain 65,536 files each. This supports an approximately 80\% - 20\% train-test split. 
More recent VirusShare chunks contain newer forms of malware that do not appear earlier in the dataset \cite{virusshare}. 
MalDICT-Behavior and MalDICT-Platform test sets contain malware added to the VirusShare corpus between July 2018 and Apr. 2023, while all of the malware in the training sets were added prior to July 2018. The first chunks were added to VirusShare in 2012, but we are aware of malware in VirusShare which was uploaded to VirusTotal in 2006 \cite{virusshare, avscan2vec}. Since new types of malware are continually being observed, 
This enables a temporal train-test split which simulates model performance on novel types of malware that do not appear in the training set. With up to nearly a five-year gap between the chunks in the training and test sets, MalDICT-Behavior and MalDICT-Platform can unveil whether a malware classifier is robust against out-of-distribution (OOD) data from a "future" time period.

The training and test sets for MalDICT-Vulnerability and MalDICT-Packer do not use a temporal split. This is because VULN and PACK tags are much less frequent, and we observed that multiple VULN and PACK tags only appear in the dataset over a short time interval. If we had used a temporal split, this would have resulted in a number of tags appearing in only the training set but not the test set or vice-versa. Instead, we used a stratified 80\% - 20\% train-test split to ensure even proportions of tags in the training and test sets.

\subsection{MalDICT Dataset Contents}
Table \ref{tab:dataset_stats} lists the number of files and number

\begin{wraptable}[4]{R}{0.5\textwidth}
\centering
\vspace*{-52pt}
\caption{Contents of MalDICT Datasets}
\label{tab:dataset_stats}
\adjustbox{max width=0.47\textwidth}{

\begin{tabular}{@{}lrrrr@{}}
\toprule
 & Total Files & Train Set & Test Set & Tags \\
\midrule

Behavior & 4,317,241 & 3,744,022 & 573,219 & 75 \\
Platform & 963,492 & 738,264 & 225,228 &  43 \\
Vulnerability & 173,886 & 136,467 & 37,419 & 128 \\    
Packer & 252,148 & 201,392 & 50,756 &  79 \\    
\bottomrule
\end{tabular}
}
\vspace*{6pt}

\end{wraptable}

\noindent  of unique tags in the four MalDICT  datasets. Due to some files occurring in multiple datasets, MalDICT includes 5,457,778 unique malicious files in total. We are releasing the file hashes and ClarAVy token rankings for each of these files. Since they are a subset of the VirusShare corpus, the corresponding malicious files can be downloaded by any malware analyst who has been granted a VirusShare login \cite{virusshare}. Furthermore, we are releasing the disarmed executable and EMBER raw metadata for each PE file in MalDICT. Files were disarmed by zeroing out the OPTIONAL\_HEADER.Subsystem and
FILE\_HEADER.Machine fields in their PE headers, which is the same method used by SOREL and MOTIF \cite{sorel,motif}.

\subsection{Sources of Bias in MalDICT}

We now survey potential sources of bias in the MalDICT datasets. To counteract the questionable accuracy of individual AV labels, we chose to only include BEH and PLAT tags for which there is consensus between at least five uncorrelated AV products \cite{botacin, mohaisen2014}. We judge this to be necessary for tag accuracy, but we are aware that it may cause a selection bias \cite{li}. Omissions or errors in AV labeling is in of itself another source of bias in our dataset \cite{botacin,mohaisen2014,mohaisen2015}. However, there is no other source which can be reasonably used as a source of malware labels at this scale \cite{kantchelian}. Finally, the methods we used for selecting files to include in MalDICT changed the tags and their distributions from what would be observed in the wild. We have already justified these design choices earlier in this section.

\section{Baseline Models}

We are releasing models trained on four MalDICT datasets. These models serve as measurements of baseline ML performance in this problem space. We selected MalConv2 and LightGBM as baseline models, since they are similar to those used by other notable datasets \cite{malconv2, lightgbm, ember, sorel, motif}.

\begin{wraptable}[6]{R}{0.55\textwidth}
\centering
\vspace*{-40pt}
\caption{MalConv2 Evaluation (Micro Avg.)}
\label{tab:malconv2-micro}

\adjustbox{max width=0.52\textwidth}{
\begin{tabular}{@{}lrrrr@{}}
\toprule
 & Behavior & Platform & Vulnerability & Packer \\
\midrule

Precision & .651 & .750 & .926 & .897 \\
Recall & .492 & .718 & .888 & .801 \\
F1-Measure & .560 & .733 & .906 & .846 \\ 
ROC-AUC & .929 & .965 & .995 & .987 \\
\bottomrule
\end{tabular}
}
\vspace*{6pt}

\end{wraptable}

\subsection{MalConv2 Baseline Model}
\label{sec:baselines}

Our first baseline model is MalConv2, a convolutional neural network that accepts raw file bytes as input \cite{malconv2}. MalConv2 was also used as a baseline model by the MOTIF dataset, and the original MalConv was used by the EMBER dataset 

\begin{wraptable}[6]{R}{0.55\textwidth}
\centering
\vspace*{-28pt}
\caption{MalConv2 Evaluation (Weighted Avg.)}
\label{tab:malconv2-weighted}
\adjustbox{max width=0.52\textwidth}{

\begin{tabular}{@{}lrrrr@{}}
\toprule
 & Behavior & Platform & Vulnerability & Packer \\
\midrule

Precision & .617 & .772 & .926 & .892 \\
Recall & .492 & .718 & .    888 & .801 \\
F1-Measure & .512 & .718 & .903 & .842 \\ 
ROC-AUC & .896 & .960 & .995 & .980 \\
\bottomrule
\end{tabular}
}
\vspace*{-12pt}
\end{wraptable} 

\noindent \cite{motif, ember, malconv, malconv2}. Our baseline MalConv2 model truncates any files greater than 1MB to 1MB in order to lessen GPU memory usage. The remaining hyperparameters were kept as the MalConv2 defaults. Then, we trained MalConv2 classifiers on the four MalDICT training sets  using eight NVIDIA RTX 6000 GPUs in parallel. The MalConv2 model for MalDICT-Behavior was trained for 33 epochs (approximately 24 hours), and the other three MalConv2 models were trained for 100 epochs each. When a file is provided as input to the baseline MalConv2 model, it outputs the probability of each tag being associated with that file. For the purposes of computing Precision, Recall, and F1-Measure, we consider an output greater than or equal to 0.5 as the threshold for predicting a tag. We used  standard definitions of Precision, Recall, and F1-Measure for these results rather than our own  definitions in Section \ref{sec:motif-val}. MalConv2 results on MalDICT are shown in Tables \ref{tab:malconv2-micro} and \ref{tab:malconv2-weighted}. MalConv2 displays good performance when classifying malware by vulnerability and packer. Performance is lower when classifying by behavior and by platform, and this is almost certainly due to the temporal train-test split present in MalDICT-Behavior and MalDICT-Platform, but not in MalDICT-Vulnerability or MalDICT-Packer.

\subsection{LightGBM Baseline Model} 

\begin{wraptable}[10]{R}{0.55\textwidth}
\centering
\vspace*{-38pt}
\caption{LightGBM OvR Evaluation (Micro Avg.)}
\label{tab:lightgbm-micro}
\adjustbox{max width=0.52\textwidth}{

\begin{tabular}{@{}lrrr@{}}
\toprule
 & Behavior & Platform &  Packer \\
\midrule

Precision & .177 & .682 & .783 \\
Recall & .555 & .953 & .948  \\
F1-Measure & .268 & .795 & .857  \\ 
ROC-AUC & .897 & .958 & .992  \\
\bottomrule
\end{tabular}
}
\vspace*{6pt}

\end{wraptable}

The EMBER feature vector format has become a de-facto standard for representing malware in the Windows Portable Executable (PE) file format \cite{ember}. Like the EMBER, SOREL, and MOTIF datasets, we use a LightGBM classifier trained on EMBER feature vectors as a baseline model. MalDICT includes malware that is not in the PE format, as well as files with corrupt or invalid PE header fields. EM-

\begin{wraptable}[5]{R}{0.55\textwidth}
\centering
\vspace*{-60pt}
\caption{LightGBM OvR Evaluation (Weighted Avg.)}
\label{tab:lightgbm-weighted}
\adjustbox{max width=0.52\textwidth}{

\begin{tabular}{@{}lrrr@{}}
\toprule
 & Behavior & Platform &  Packer \\
\midrule

Precision & .363 & .889 & .844 \\
Recall & .555 & .953 & .948  \\
F1-Measure & .385 & .911 & .884  \\ 
ROC-AUC & .805 & .955 & .991  \\
\bottomrule
\end{tabular}
}
\vspace*{6pt}

\end{wraptable}

\noindent  BER vectors for these files could not be computed, so they were excluded from this experiment. Nearly all of the malware in MalDICT-Vulnerability are malicious scripts rather than PE files, so we did not train a LightGBM model on this data. In the remaining three MalDICT datasets, there were a small number of tags which contained little to no PE files, and they were also excluded. Since this is a multiclass, multilabel problem, we trained one-versus-Rest (OvR) LightGBM classifiers on each tag for 100 iterations each. Results are displayed in Tables \ref{tab:lightgbm-micro} and \ref{tab:lightgbm-weighted}.

The LightGBM classifier performed well on MalDICT-Platform and MalDICT-Packer, but was extremely poor at classifying malware in MalDICT-Behavior. MalConv2 performance on MalDICT-Behavior was substandard as well, but not to such an extent. MalConv2 and LightGBM both performed worse on MalDICT-Behavior than the SOREL dataset's feed-forward neural network (FFNN) baseline classifier, which achieved ROC-AUC scores above 0.97 for all 11 of its behavioral tags \cite{sorel}. We believe that the temporal train-test split and the increased number of tags in MalDICT-Behavior result in a more difficult classification problem than the SOREL dataset offers. Recall that the most recent malware in MalDICT-Behavior's test set was added in April 2023, while the most recent malware in its training set was added in July 2018. This makes MalDICT-Behavior a true test on a malware classifier's OOD performance. If a model performs well on this benchmark, practitioners can be assured that the model can generalize to malicious attributes that are present in malware far into the "future".

\section{Conclusion}
\label{sec:conclusion}

To our knowledge, MalDICT includes the first public malware datasets labeled according to platform, vulnerability, and packer. It also includes the most diverse public dataset of malware labeled by behavior, containing over 4.3 million malicious files and 75 distinct behavioral tags. We are releasing the file hashes and tags for the nearly 5.5 million malicious files in MalDICT. We are also releasing the EMBER raw features and disarmed executable files for all of the malware in MalDICT with the PE format. All of the malware in MalDICT can be obtained by researchers who have been granted to the VirusShare corpus.

Additionally, we are publishing ClarAVy, the tool that was used to accurately tag the malware in MalDICT. With support for 90 different AV products and 882 different AV label formats, ClarAVy offers more comprehensive parsing than any other AV-based  malware tagging tool. ClarAVy can extract tags from tens of millions of AV scan reports, which can then be used to train production malware classifiers. 

Our baseline classifier results indicate that there is significant room for improvement on all four tasks that MalDICT supports, especially malware behavior classification. The development of a classifier with strong performance on MalDICT-Behavior would represent a major success towards resisting concept drift over years of malware evolution. It is our hope that these contributions will facilitate and encourage further study of atypical malware classification tasks, fostering improved understanding and defense.

\bibliography{sample-base}

\end{document}